\documentclass[a4paper]{jpconf}
\usepackage[utf8]{inputenc}
\usepackage{amsmath}
\usepackage{amsfonts}
\usepackage{amssymb}
\usepackage{graphicx}
\usepackage{siunitx}
\usepackage{xcolor}
\usepackage{hyperref}

\begin{document}
\title{A Proton Computed Tomography Demonstrator for Stopping Power Measurements}
		\author{Felix Ulrich-Pur$^1$, Thomas Bergauer$^1$, Alexander Burker$^2$, Albert Hirtl$^2$, Christian Irmler$^1$, Stefanie Kaser$^1$, Florian Pitters$^1$}
        \address{$^1$ Austrian Academy of Sciences, Institute of High Energy Physics (HEPHY),
        Nikolsdorfer Gasse 18, 1050 Wien, Austria}
		\address{$^2$ TU Wien, Atominstitut, Stadionallee 2, 1020 Wien, Austria}
		\ead{felix.ulrich-pur@oeaw.ac.at}
\begin{abstract}
Particle therapy is an established method to treat deep-seated tumours using accelerator-produced ion beams. For treatment planning, the precise knowledge of the relative stopping power (RSP) within the patient is vital. Conversion errors from x-ray computed tomography (CT) measurements to RSP introduce uncertainties in the applied dose distribution. Using a proton computed tomography (pCT) system to measure the SP directly could potentially increase the accuracy of treatment planning. A pCT demonstrator, consisting of double-sided silicon strip detectors (DSSD) as tracker and plastic scintillator slabs coupled to silicon photomultipliers (SiPM) as a range telescope, was developed. After a significant hardware upgrade of the range telescope, a 3D tomogram of an aluminium stair phantom was recorded at the MedAustron facility in Wiener Neustadt, Austria. In total, 80 projections with $\SI{6.5e5}{}$ primary events were acquired and used for the reconstruction of the RSP distribution in the phantom. After applying a straight-line approximation for the particle path inside the phantom, the most probable value (MPV) of the RSP distribution could be measured with an accuracy of $\SI{0.59}{\percent}$. The RSP resolution inside the phantom was only $\SI{9.3}{\percent}$ due to a limited amount of projections and measured events per projection.
\end{abstract}
\section{Introduction}
Due to the strongly localized energy deposition of heavy charged ions in matter, protons are well suited for treating tumours close to vital organs while reducing exposure to normal tissues \cite{Miller1995}. However, the deposited dose within the patient strongly depends on the initial beam energy as well as the stopping power (SP) of the traversed material, which is usually expressed relative to water (RSP). Therefore, an accurate RSP map of the patient is vital for an optimal treatment plan. Current treatment plans are based on x-ray computed tomography (CT), where the obtained Hounsfield units (HU) have to be converted into RSP. The extrapolation from HU to RSP introduces conversion errors, leading to uncertainties in the applied dose distribution \cite{Schaffner1998}. Alternatively, the RSP distribution could be measured directly with a proton computed tomography (pCT) system to potentially increase the treatment planning accuracy \cite{Dedes2019}.
\section{pCT setup}
\begin{figure}[h]
\begin{minipage}{25pc}
\centering
\includegraphics[width=25pc]{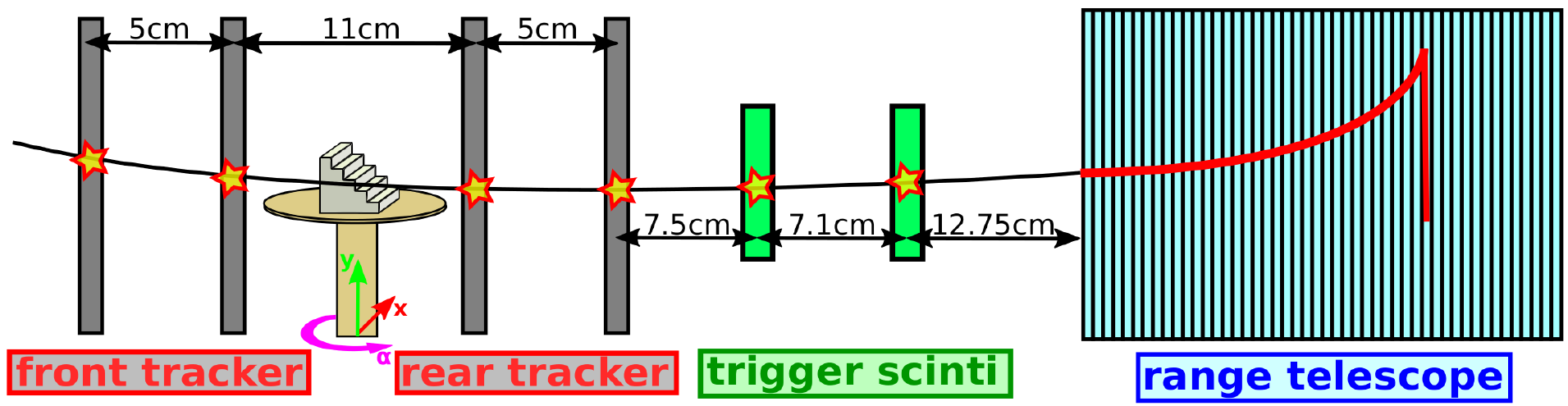}
\caption{\label{fig:setupsketch}Schematic overview of the pCT demonstrator. In the final experimental set-up, the aluminium phantom was  mounted upside-down on a rotating table due to geometric constraints.}
\end{minipage}\hspace{2pc}%
\begin{minipage}{9pc}
\centering
\includegraphics[width=9pc]{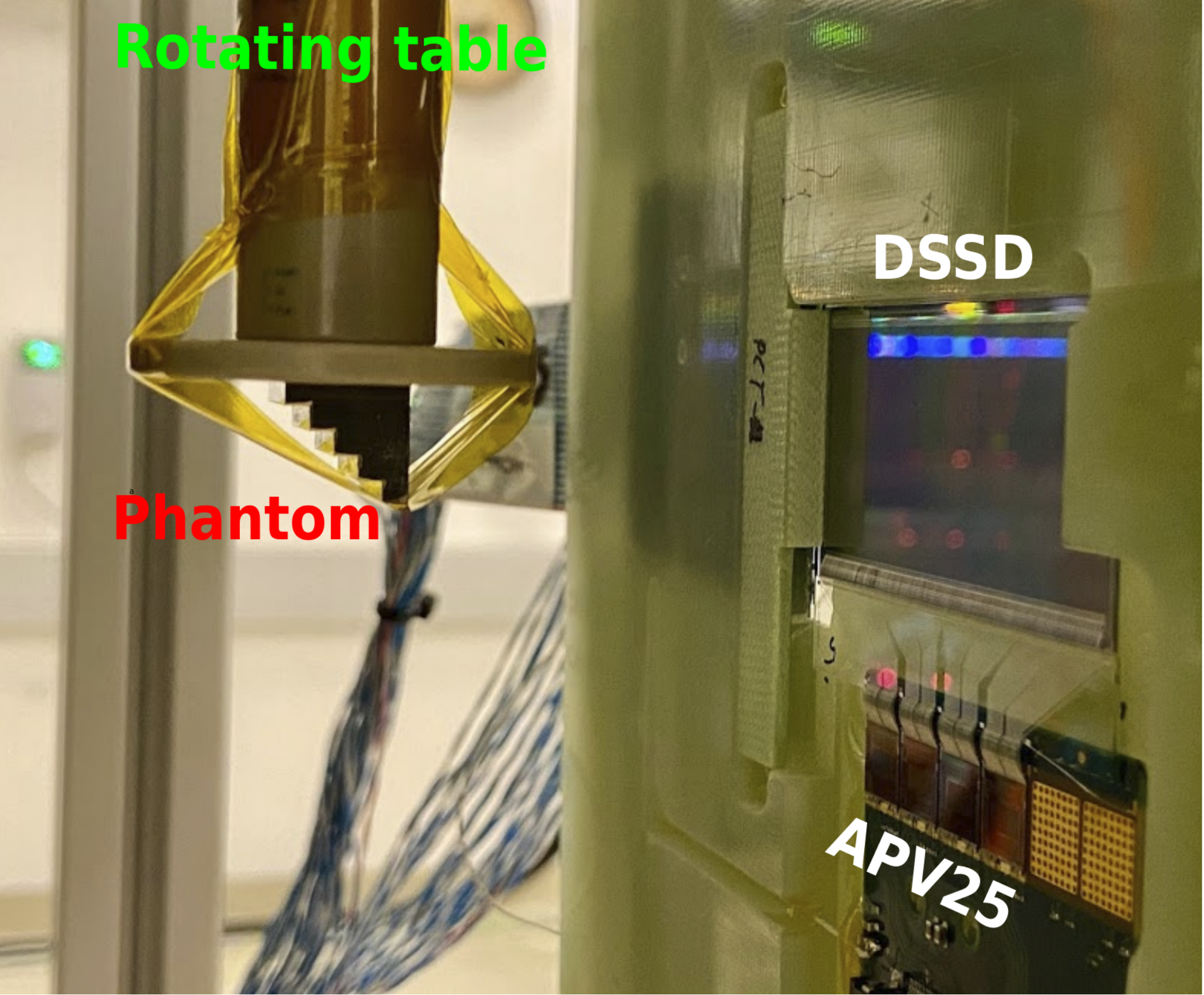}
\caption{\label{fig:trackerandphantom}Al phantom next to the n-side ($y$-coordinate) of the DSSD.}
\end{minipage} 
\end{figure}

In a standard pCT system \cite{schulte2004}, the particle path \cite{Collins_Fekete_2017} and the deposited energy of a particle passing through the patient are measured at different incident angles to reconstruct the 3D map of the RSP in the patient. As a first step to establish pCT in our group, a pCT demonstrator was built from existing hardware before building a full clinical pCT system. As depicted in Figures \ref{fig:setupsketch} and \ref{fig:trackerandphantom}, instead of a patient, a $\SI{1}{cm^3}$ Aluminium cube phantom with five steps was used to test the performance of the pCT demonstrator. The phantom was mounted on a rotating table and placed between two tracking detector pairs which were used to measure the position and direction of the proton entering and leaving the phantom. Due to the small size of the aluminium cube in contrast to a full-size patient, a straight-line approximation \cite{KASERTIGRE2021} was used to estimate the particle path through the phantom. In order to estimate the deposited energy inside the phantom, a residual range telescope was placed downstream of the second tracker pair. To synchronize the range telescope with the tracker, the AIDA2020 trigger and logic unit (TLU)\cite{Baesso_2019} was used with two $\SI{1}{}\times\SI{5}{}\times\SI{5}{cm^3}$ plastic scintillators as trigger.  

\subsection{Tracker}
In contrast to a previous version of this pCT demonstrator \cite{ULRICHPUR2020Hiroshima}, four instead of six n-substrate based double-sided silicon strip detectors (DSSDs) were used for the tracker. The sensors are $\SI{300}{\mu m}$ thick and have an active area of $\SI{2.56}{}\times \SI{5.12}{cm^2}$. Each side of the DSSDs features 512 AC-coupled strips with a pitch of $\SI{100}{\mu m}$ on the n-side ($x$-coordinate) and $\SI{50}{\mu m}$ on the p-side ($y$-coordinate).   The strips on each side are connected to 4 APV25 \cite{APV25} chips (Figure \ref{fig:trackerandphantom}) which are then read out by a
VME-based flash ADC (FADC) system \cite{ULRICHPUR2020Hiroshima}, initially developed for Belle II \cite{BelleIISVDReadout}. In order to increase the maximum data acquisition (DAQ) rate of the tracker, the old VME-based data transfer from the FADC boards to the PC was replaced by a Gigabit Ethernet (GbE) readout system, increasing the DAQ rate from $\SI{500}{Hz}$ to $\SI{30}{kHz}$.
\subsection{Range telescope}
In pCT, the energy loss inside the phantom is used to calculate the water equivalent thickness (WET \cite{Zhang2009}) of the traversed path inside the phantom. Therefore the PRR30 range telescope, formerly developed by TERA \cite{Bucciantonio}, was used to directly measure the WET. The residual range in the telescope was measured with 42 plastic scintillator slabs with a thickness of $\SI{3}{mm}$ and an active area of $\SI{30}{}\times\SI{30}{cm^2}$. Each slab is connected to $\SI{1}{mm^2}$ SiPMs with 400 pixels via a wave length shifting fiber. The signal of each SiPM is forwarded to a FPGA-based ASIC, after being digitized with a 12-bit ADC. The FPGA is then readout via USB with a maximum DAQ rate of $\SI{16e3}{events/s}$. Since the original residual range telescope suffered from severe voltage instabilities \cite{ULRICHPUR2020Hiroshima},\cite{dissbuccantonio}, the mainboard was completely redesigned. In addition, the LabView based read-out software was completely replaced by a custom C++ based DAQ software with automated calibration procedures \cite{benjaminma}. 
\subsubsection{Range telescope calibration}
\begin{figure}[h]
\begin{minipage}{18pc}
\includegraphics[width=18pc]{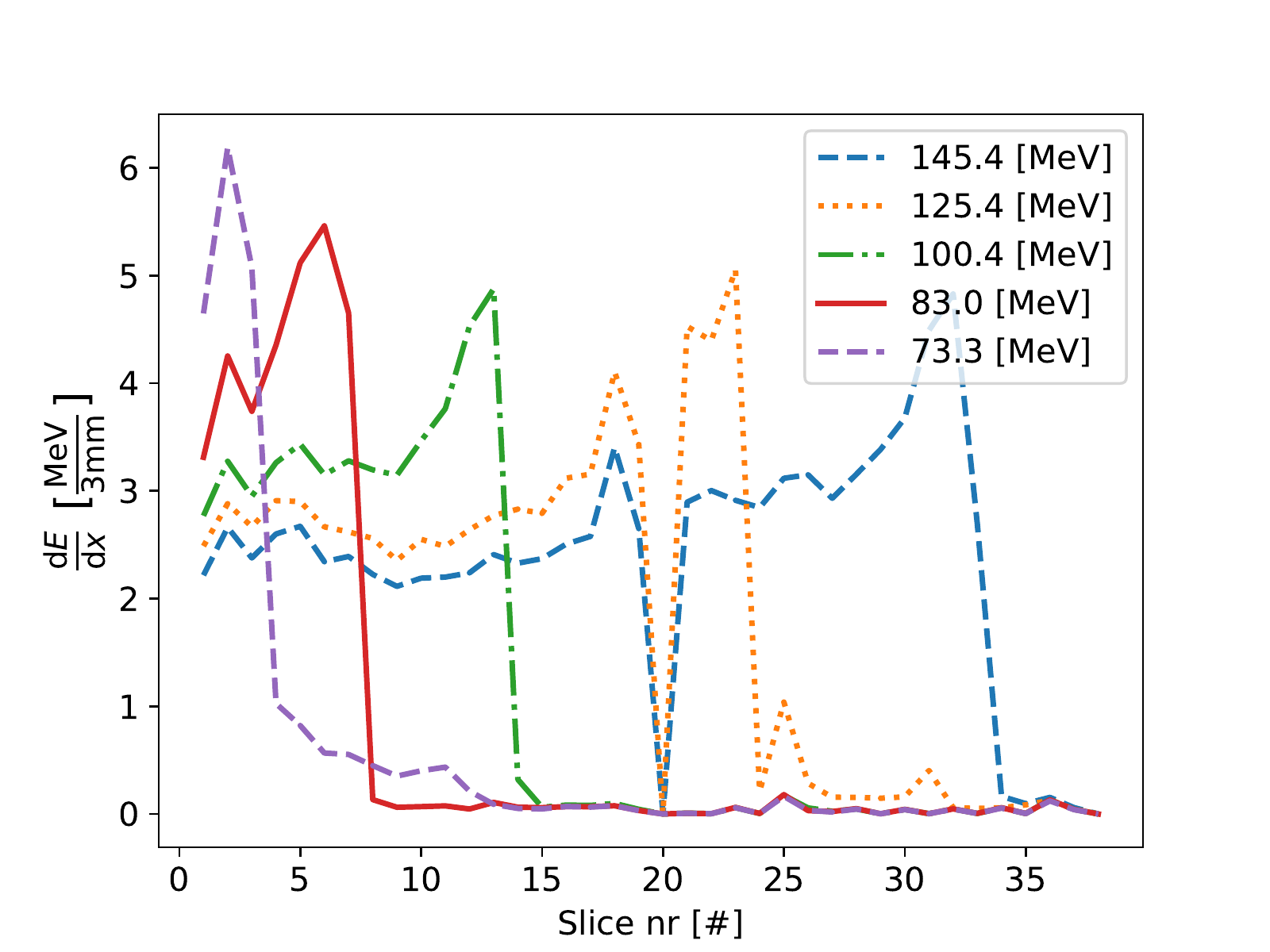}
\caption{\label{fig:braggpeaks}Bragg peaks}
\end{minipage}\hspace{2pc}%
\begin{minipage}{18pc}
\includegraphics[width=18pc]{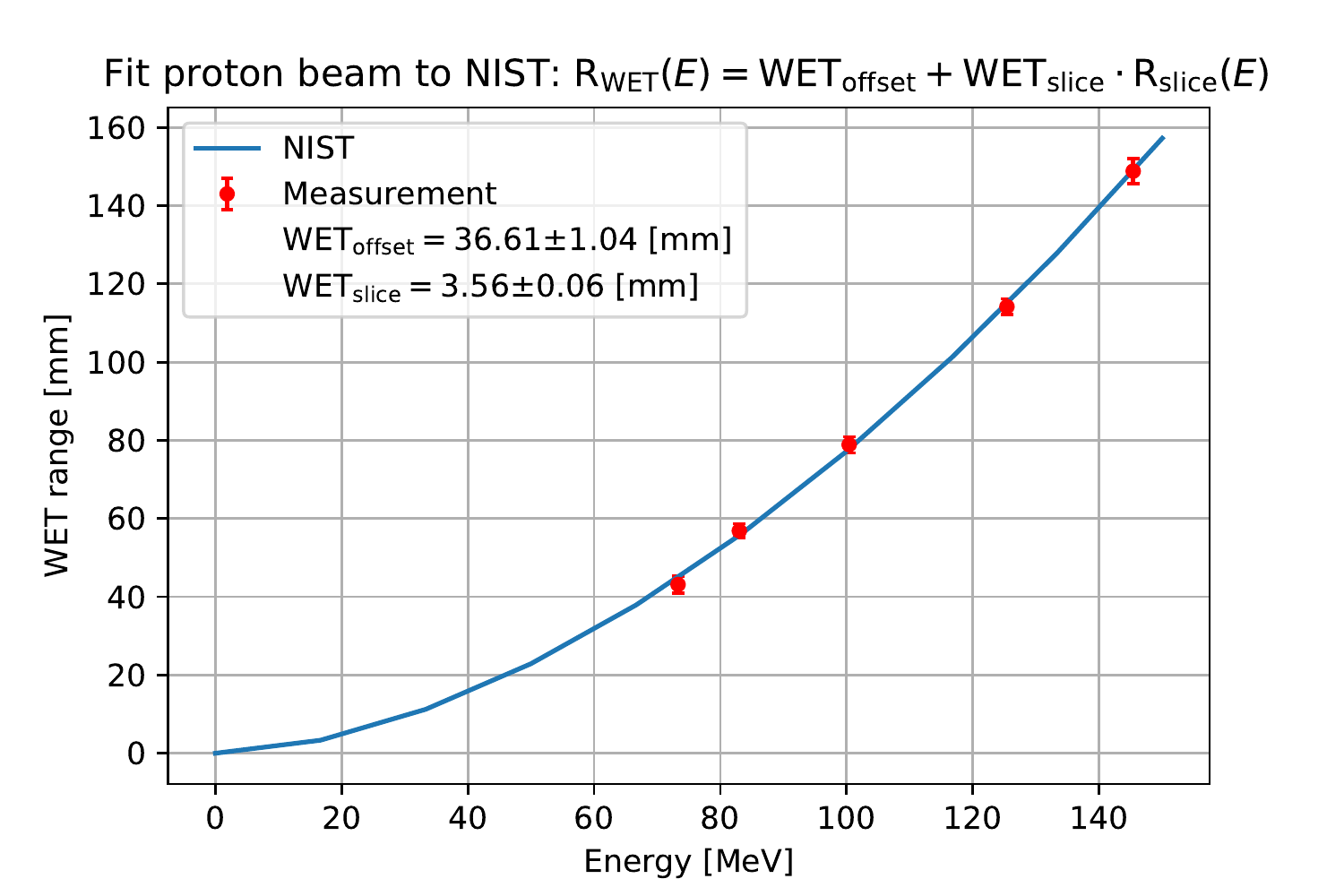}
\caption{\label{fig:teracalib}TERA calibration curve}
\end{minipage} 
\end{figure}

\mbox{}\\
Prior to the pCT measurement, the upgraded range telescope had to be calibrated. First, the working point of the SiPMs was optimized by adjusting the gain of the SiPMs. For that purpose the range telescope was exposed to $\SI{800}{MeV}$ protons at MedAustron. The calibration factors, to convert the acquired ADC counts into deposited energy per slice, were then obtained by comparing the most probable value (MPV) of the measured energy spectrum per slice to a Geant4 \cite{AGOSTINELLI2003250} simulation. Due to the small number of pixel per SiPM (400) and low ADC resolution (12-bit), the signal in the SiPMs always saturated in the Bragg peak region (Figure \ref{fig:braggpeaks}). Therefore the TERA range telescope was only used as a range telescope and not as a sampling calorimeter. After calibrating each slice independently, the range telescope was calibrated by measuring the WET of the trigger scintillators as well as the mean WET per scintillator slice. This was done by measuring the residual ranges for different proton energies (Figure \ref{fig:braggpeaks}) and comparing them to the theoretical ranges in water obtained from the NIST database \cite{berger2017stopping} (Figure \ref{fig:teracalib}). The residual range in the telescope was defined as last slice over threshold and first slice under threshold to compensate for signal fluctuations in single slices.

\section{Imaging results}
The performance of the pCT demonstrator was then tested in a testbeam in the research beam line of MedAustron. $\SI{100.4}{MeV}$ protons with reduced beam intensities ($\approx \SI{3e5}{particles/s}$) \cite{ulrichpur2021commissioning} were used for the measurement. In total, 80 proton computed radiographs (forward projections) were recorded with angles ranging from $\SI{0}{}$ to $\SI{360}{^{\circ}}$. Not all angles were equidistant since the step size of the motor of the rotational table was limited to $\SI{1.8}{^{\circ}}$. Per rotational angle, $\SI{2.5e6}{}$ primary events were acquired with a mean acquisition time of $\approx \SI{24}{min}$. However, due to synchronization and tracking inefficiencies, only $\approx \SI{6.5e5}{}$ synchronized events could be used per projection.
\subsection{Forward projections}
\begin{figure}[h]
\begin{minipage}{12pc}
\includegraphics[width=12pc]{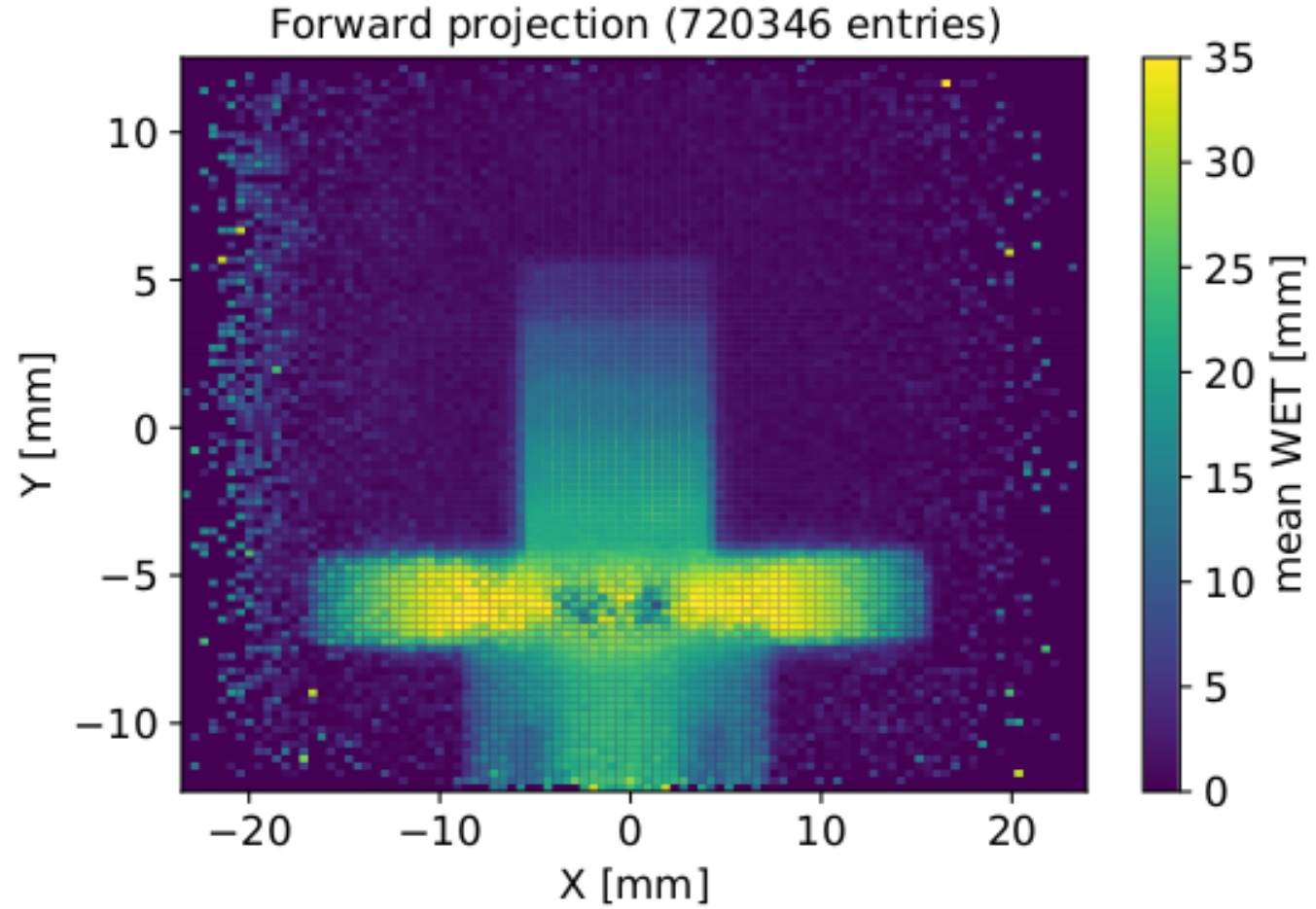}
\caption{\label{fig:0deg}Proton computed radiography at $\SI{0}{^{\circ}}$.}
\end{minipage}\hspace{1pc}%
\begin{minipage}{12pc}
\includegraphics[width=12pc]{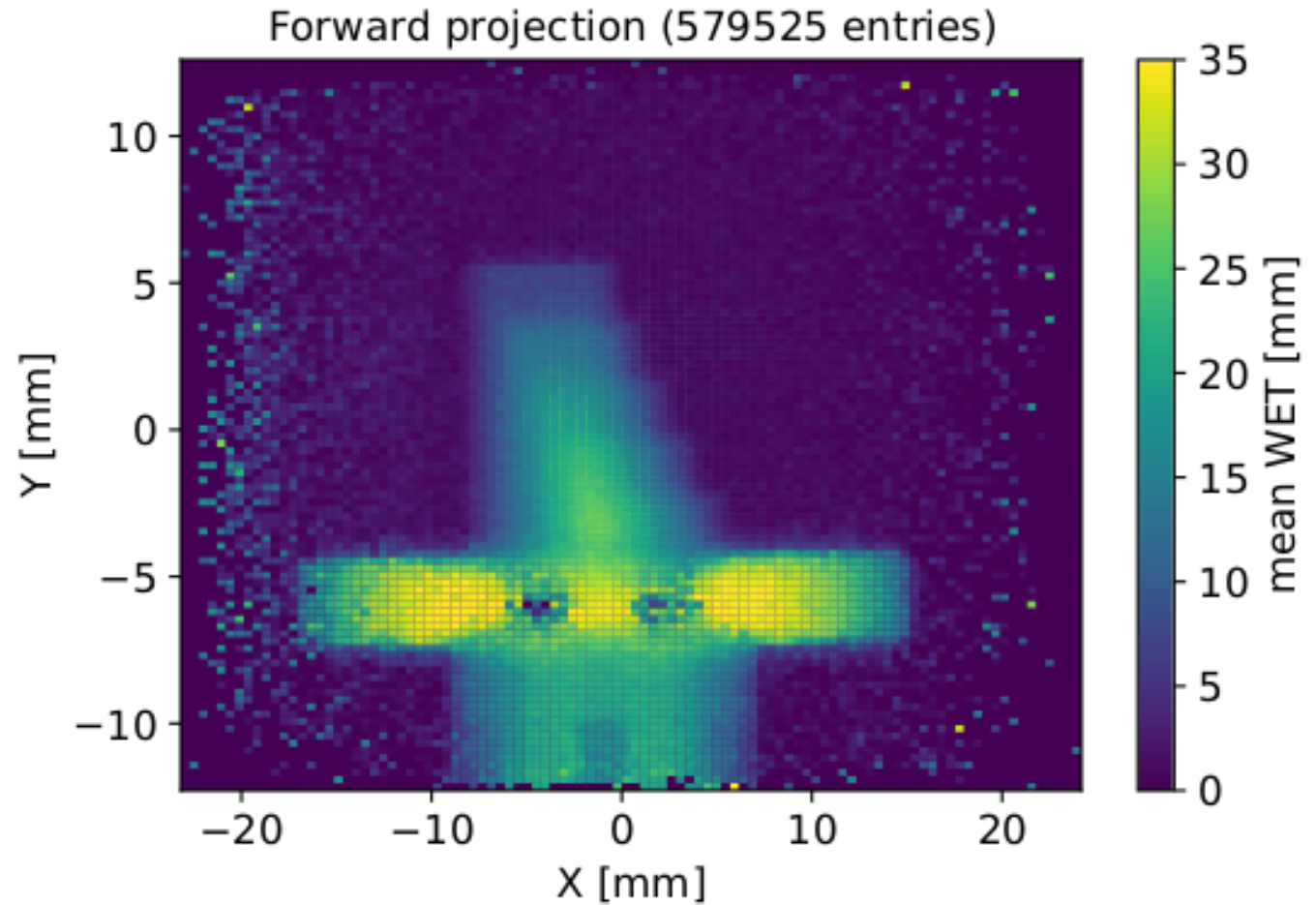}
\caption{\label{fig:45deg}Proton computed radiography at $\SI{45}{^{\circ}}$.}
\end{minipage} \hspace{1pc}%
\begin{minipage}{12pc}
\includegraphics[width=12pc]{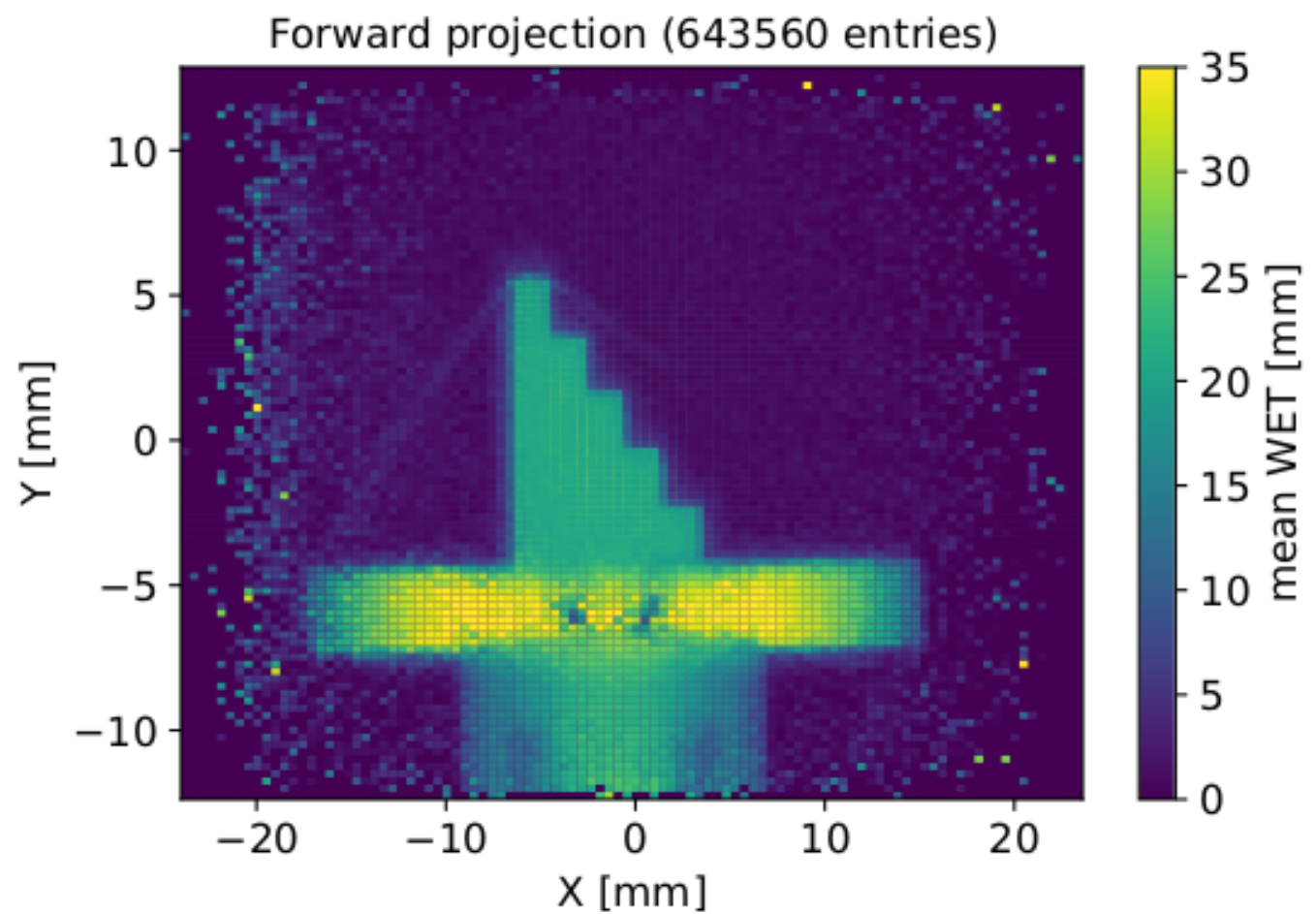}
\caption{\label{fig:90deg}Proton computed radiography at $\SI{90}{^{\circ}}$}
\end{minipage}
\end{figure}

For each forward projection, the measured WET per event was projected onto a plane perpendicular to the beam direction using the tracks of the upstream and downstream tracker. In addition to the standard $3 \sigma$-cuts to eliminate large-angle scattering events \cite{Schulte2008}, additional positional cuts, as described in \cite{KASERTIGRE2021}, were used to filter non-straight-line events.  Figures \ref{fig:0deg}-\ref{fig:90deg} show three forward projections at $\SI{0}{^{\circ}}$,$\SI{45}{^{\circ}}$ and $\SI{90}{^{\circ}}$. \if Also the Kapton tape, which was used to tape the phantom onto the rotating table (Figure \ref{fig:trackerandphantom}) is faintly visible in Figure \ref{fig:90deg}.\fi

\subsection{Reconstruction}
\begin{figure}[h]
\begin{minipage}{18pc}
\centering
\includegraphics[width=14pc]{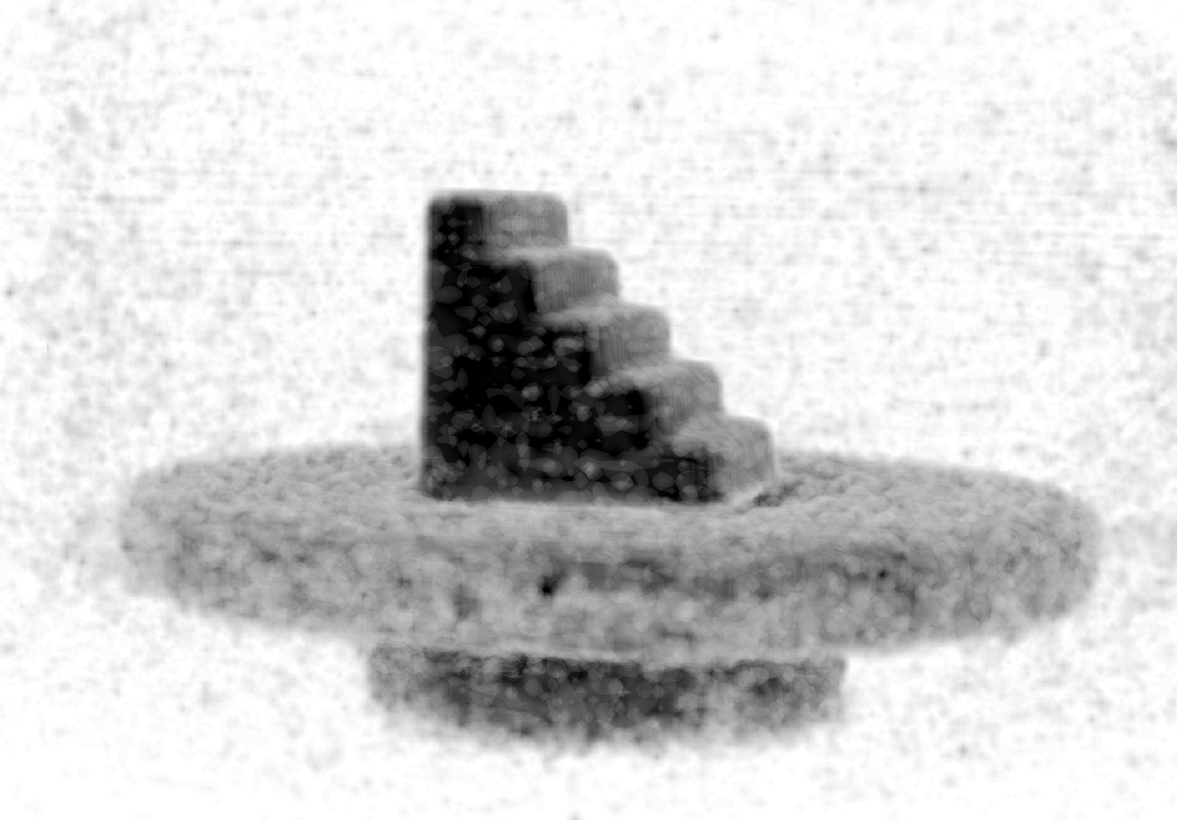}
\caption{\label{fig:3drsp}3D view of the reconstructed aluminum phantom mounted on the rotating table.}
\end{minipage}\hspace{2pc}%
\begin{minipage}{18pc}
\vspace{-0.6cm}
\centering
\includegraphics[width=17pc]{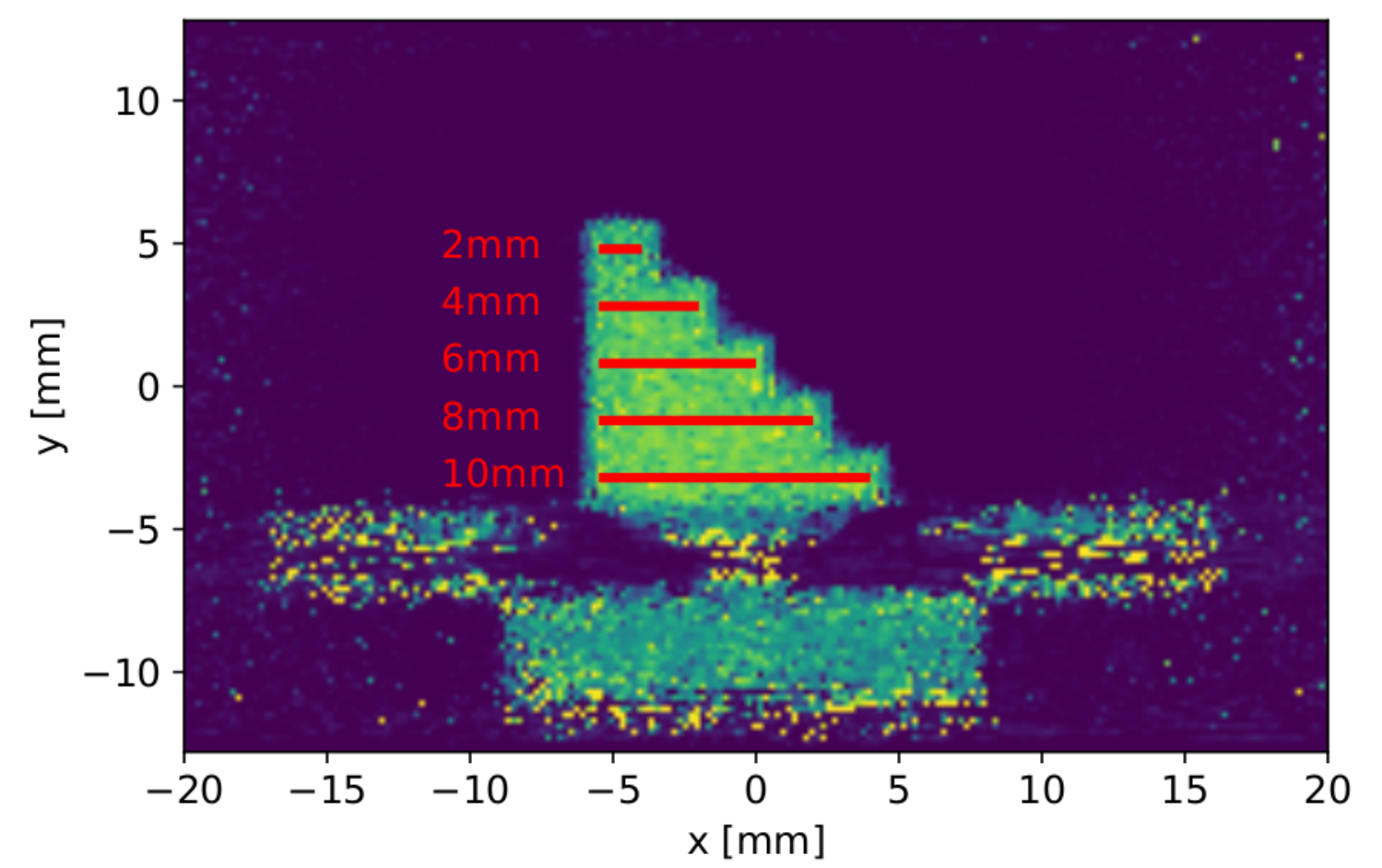}
\vspace{-0.2cm}
\caption{\label{fig:centralslice}Central slice of the recorded RSP tomogram.}
\end{minipage} 
\end{figure}

\begin{figure}[h]
\begin{minipage}{18pc}
\centering
\includegraphics[width=18pc]{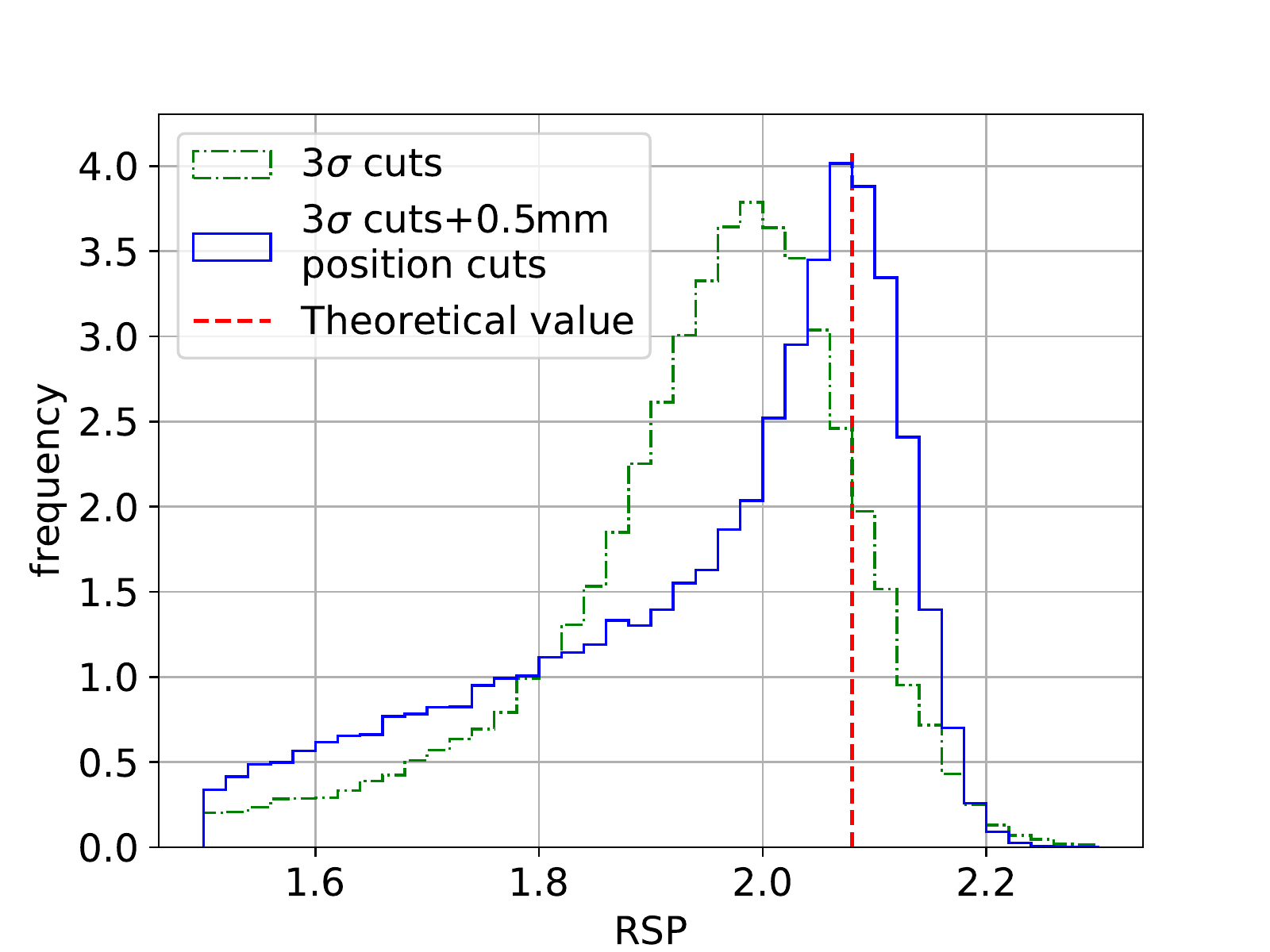}
\caption{\label{fig:rspdistr}RSP distribution measured in the Al phantom for different data cuts.}
\end{minipage}\hspace{2pc}%
\begin{minipage}{18pc}
\vspace{0.5cm}
\centering
\includegraphics[width=18pc]{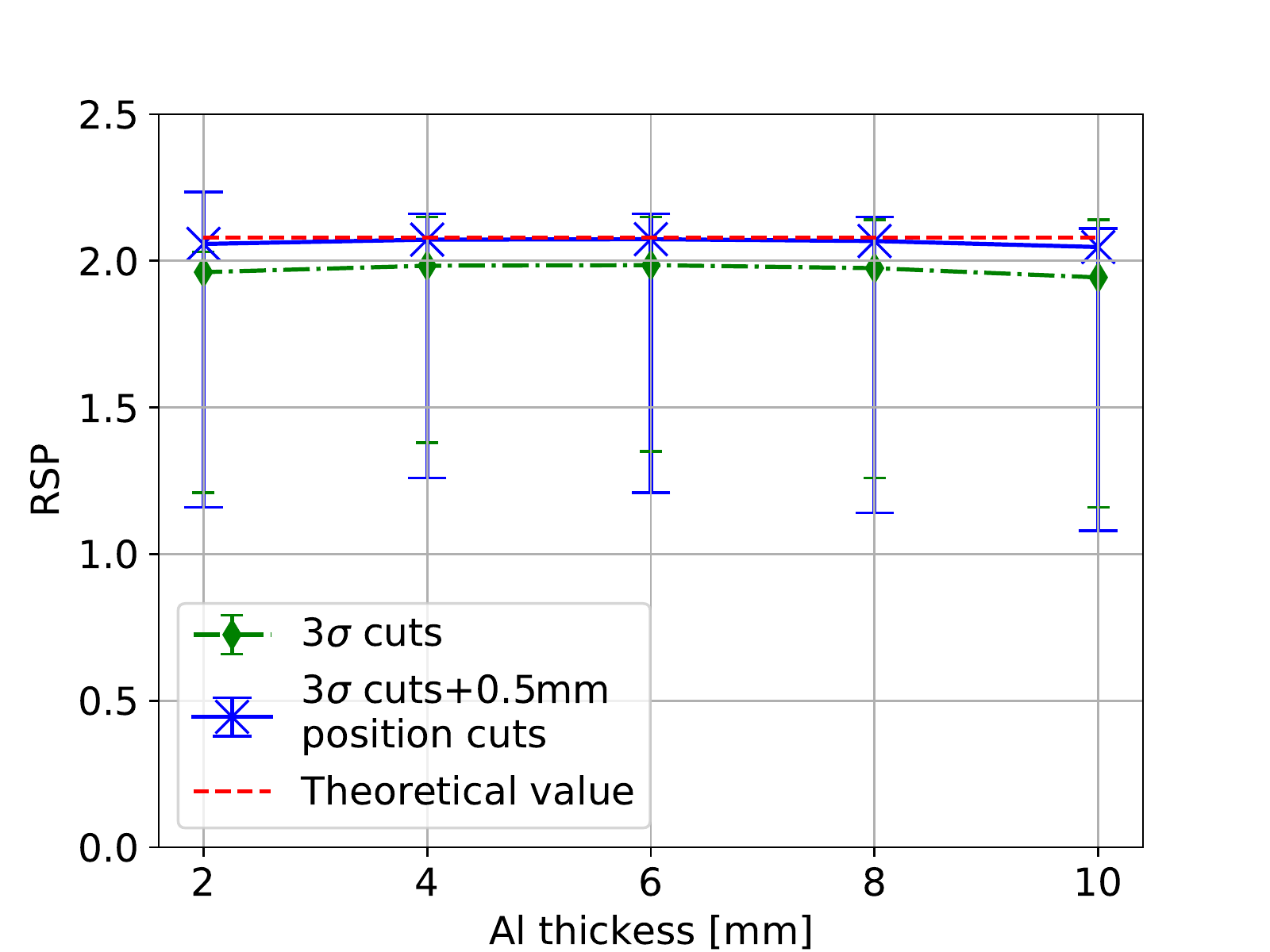}
\vspace{-0.5cm}
\caption{\label{fig:rspperstep}MPV of the RSP distribution measured in each step of the Al phantom. Also the $\SI{2.5}{}$ and $\SI{97.5}{\percent}$ quantiles are shown.}
\end{minipage} 
\end{figure}

To reconstruct the 3D RSP map inside the aluminium phantom, the reconstruction framework TIGRE (Tomographic Iterative GPU-based REconstruction toolbox) \cite{TIGRE} was used since it is able to reconstruct images with non-equidistant projection angles and a limited amount of projections. The reconstruction was performed using CUDA on an Nvidia GeForce GTX 1080 TI with the iterative algorithm OS-SART \cite{Penfold2015} with ten iterations. A 3D view of the resulting image, with a voxel side length of $\SI{0.2}{mm}$, is shown in Figure \ref{fig:3drsp}. In order to determine the RSP accuracy and precision, the RSP was acquired at different regions of interest (ROIs) inside the reconstructed RSP map (Figure \ref{fig:centralslice}). Figure \ref{fig:rspdistr} shows the RSP distribution of all ROIs combined for different data cuts. After applying additional position cuts to filter non-straight-line events ($\approx \SI{66}{\percent}$ for $\SI{0.5}{mm}$ cuts), the relative error of the MPV of the RSP distribution was lowered down to $\SI{0.59}{\percent}$. Still, a tail towards lower RSP values could be observed. More statistics would be required to apply stronger position cuts to filter the remaining non-straight-line events. Similar results were found in the individual ROIs (Figure \ref{fig:rspperstep}). The relative error of the MPV of the RSP distribution measured at each step of the phantom ranged from $\SI{0.28}{}-\SI{1.56}{\percent}$.
However, due to the limited amount of projections and events per projection, a total RSP resolution $\frac{\sigma_{\mathrm{RSP}}}{\mu_{\mathrm{RSP}}}$ of $\SI{9.3}{\percent}$ was obtained.


\section{Discussion and Outlook}
After upgrading and re-calibrating the TERA range telescope, the pCT demonstrator could be used to acquire a pCT scan of an aluminium cube phantom in the research beam line of MedAustron. The RSP accuracy inside the phantom could be decreased to $\SI{0.56}{\percent}$ after applying additional position cuts to filter non-straight-line events. However, due to the limited rate capability of the demonstrator and the limited size of the DSSDs, a different detector design should be used for a clinical pCT system, which is currently under investigation.  
\section{Acknowledgments}
The authors would like to thank A. Bauer, W. Brandner, K. Fischer, S. Schultschik, H. Steininger, R. Thalmeier and H. Yin for their contributions to the construction of the pCT demonstrator. This project has received funding from the Austrian Research Promotion Agency (FFG), grant numbers 869878 and 875854.
\section*{References}
\bibliography{tippPaperUlrich-Pur2021}

\providecommand{\newblock}{}
\begin{thebibliography}{10}
\expandafter\ifx\csname url\endcsname\relax
  \def\url#1{{\tt #1}}\fi
\expandafter\ifx\csname urlprefix\endcsname\relax\def\urlprefix{URL }\fi
\providecommand{\eprint}[2][]{\url{#2}}

\bibitem{Miller1995}
Miller D~W 1995 {\em Medical Physics\/} {\bf 22} 1943--1954

\bibitem{Schaffner1998}
Schaffner B {\em et~al.\/} 1998 {\em Physics in Medicine and Biology\/} {\bf
  43} 1579--1592

\bibitem{Dedes2019}
Dedes G {\em et~al.\/} 2019 {\em Physics in Medicine {\&} Biology\/} {\bf 64}
  165002

\bibitem{schulte2004}
Schulte R {\em et~al.\/} 2004 {\em IEEE Transactions on Nuclear Science\/} {\bf
  51} 866--872

\bibitem{Collins_Fekete_2017}
Collins-Fekete C~A {\em et~al.\/} 2017 {\em Physics in Medicine and Biology\/}
  {\bf 62} 1777--1790

\bibitem{KASERTIGRE2021}
Kaser S {\em et~al.\/} 2021 {\em Physica Medica\/} {\bf 84} 56--64 ISSN
  1120-1797

\bibitem{Baesso_2019}
Baesso P {\em et~al.\/} 2019 {\em Journal of Instrumentation\/} {\bf 14}
  P09019--P09019

\bibitem{ULRICHPUR2020Hiroshima}
Ulrich-Pur F {\em et~al.\/} 2020 {\em Nucl. Instr. and Meth. A\/} {\bf 978}
  164407

\bibitem{APV25}
French M {\em et~al.\/} 2001 {\em Nucl. Instr. and Meth. A\/} {\bf 466} 359 --
  365 ISSN 0168-9002

\bibitem{BelleIISVDReadout}
Thalmeier R {\em et~al.\/} 2017 {\em Nucl. Instr. and Meth. A\/} {\bf 845} 633
  -- 638 ISSN 0168-9002

\bibitem{Zhang2009}
Zhang R {\em et~al.\/} 2009 {\em Physics in Medicine and Biology\/} {\bf 54}
  1383--1395

\bibitem{Bucciantonio}
Bucciantonio M {\em et~al.\/} 2013 {\em Nucl. Instr. and Meth. A\/} {\bf 732}
  564 -- 567 ISSN 0168-9002

\bibitem{dissbuccantonio}
Bucciantonio M 2015 {\em Development of an advanced {P}roton {R}ange
  {R}adiography system for hadrontherapy\/} Ph.D. thesis
  \href{https://swisscovery.slsp.ch/permalink/41SLSP_NETWORK/1ufb5t2/alma991084316869705501}{Philosophisch-naturwissenschaftliche
  Fakult\"{a}t der Universit\"{a}t Bern}

\bibitem{benjaminma}
Huber B 2020 {\em Development and assessment of a calorimeter data acquisition
  and evaluation software for ionizing particles\/} Master's thesis \href{
  https://permalink.catalogplus.tuwien.at/AC16081590}{TU Wien}

\bibitem{AGOSTINELLI2003250}
Agostinelli S {\em et~al.\/} 2003 {\em Nucl. Instr. and Meth. A\/} {\bf 506}
  250--303 ISSN 0168-9002

\bibitem{berger2017stopping}
Berger M {\em et~al.\/} 2017 Stopping-power and range tables for electrons,
  protons, and helium ions, {NIST} {S}tandard {R}eference {D}atabase 124

\bibitem{ulrichpur2021commissioning}
Ulrich-Pur F {\em et~al.\/} 2021 {\em Nucl. Instr. and Meth. A\/}  165570

\bibitem{Schulte2008}
Schulte R~W {\em et~al.\/} 2008 {\em Medical Physics\/} {\bf 35} 4849--4856

\bibitem{TIGRE}
Biguri A {\em et~al.\/} 2016 {\em Biomedical Physics {\&} Engineering
  Express\/} {\bf 2} 055010

\bibitem{Penfold2015}
Penfold S {\em et~al.\/} 2015 {\em Sensing and Imaging\/} {\bf 16}

\end{thebibliography}
\end{document}